\documentclass[10pt,british]{article}
\usepackage{mathptmx}
\usepackage[T1]{fontenc}
\usepackage[latin9]{luainputenc}
\usepackage{babel}
\usepackage{mathrsfs}
\usepackage{amsmath}
\usepackage{amssymb}
\usepackage{graphicx}
\usepackage{setspace}
\usepackage[pdfusetitle,
 bookmarks=true,bookmarksnumbered=false,bookmarksopen=false,
 breaklinks=false,pdfborder={0 0 1},backref=false,colorlinks=false]
 {hyperref}

\makeatletter
\newcommand{\lyxaddress}[1]{
	\par {\raggedright #1
	\vspace{1.4em}
	\noindent\par}
}

\makeatother

\begin{document}
\title{Universal scaling limits for spin networks via martingale methods }
\author{Simone Franchini}
\date{~}
\maketitle

\lyxaddress{\begin{center}
\textit{Sapienza Università di Roma, Piazza Aldo Moro 1, 00185 Roma,
Italy}
\par\end{center}}
\begin{abstract}
We use simple martingale methods to construct a large deviation theory
of spin systems with pairwise interactions. As an application, we
show that the fully connected case obeys a universal scaling limit
that is just a product of magnetisation eigenstates.

~


~
\end{abstract}
In \cite{FranchiniSPA2021,FranchiniSPA2023,FranchiniRSBwR2023} we
elaborated a Large--Deviations theory for disordered systems that
allows to study any spin system by means of standard combinatorial
techniques, including the celebrated Sherrington--Kirckpatrick (SK)
model, a mean field paradigm for complex systems \cite{RSB}. Here,
we used the same methods to show that any fully connected model with
any pairwise interactions, including the random models (i.e., models
with random parameters, like the SK model), obeys a universal scaling
limit. We actually construct the scaling limit of the free energy
density in terms of the magnetisations eigenstates, i.e., spin states
with fixed average magnetisation, in the Eq. (\ref{eq:Bound}). 

The main passages of our analysis were: to introduce a sequence of
sub--systems of increasing sizes by partitioning the vertex set into
a number of disjoint sub--sets (that in the references \cite{FranchiniSPA2021,FranchiniSPA2023,FranchiniRSBwR2023}
we called ``Layers'') and then by re--joining these up to a certain
point of that sequence: this is explained in the Section \ref{sec:The-general-model}.
Then, we reinterpret them as a series of  Markov blankets, \cite{Judea_Pearl,Friston},
and identify their Hamiltonian sequence in Section \ref{sec:Blankets-(or-Layers)}.
We find that each blanket can be further destructured in two general
contributions: the first term, ``core'' contribution, is the energy
from the interactions inside the blanket itself, while the second,
the ``interface'', is due to the interactions between the spins
of the blanket and the rest of the network. Formally, it is the scalar
product between the spins of the blanket and local fields that depends
on the previous blankets. As we shall see this is a general theory
and, up to this point, it could be used to deal with any spin system.
Then, in Section \ref{sec:Fully-connected-models}, we show that in
a fully connected model if the partition is refined enough the core
contribution is small and the thermodynamics is dominated by the interfaces.
Finally, in the Section \ref{subsec:General-theory-of}, we show that
the Hamiltonian with the cores removed (IO model, \cite{FranchiniRSBwR2023})
obeys a universal scaling limit that can be expressed in terms of
the magnetisation eigenstates like in the SK model \cite{FranchiniRSBwR2023,MezVir}. 

Apart from providing characterisation for the scaling limit of any
fully connected spin network in terms of magnetisation eigenstates,
an important goal of this paper is to provide a simple presentation
of the method introduced in \cite{FranchiniSPA2021,FranchiniSPA2023,FranchiniRSBwR2023}
that is readable by people working in the fields of Bayesian Inference,
Machine Learning \cite{Judea_Pearl}, and Active Inference \cite{Friston,Zeidman,Heins,Bardella}.
Given this purpose, here we avoid the details of the underlying kernel
theory (KT) \cite{FranchiniRSBwR2023,Bardella,BFPF,Franchini2017,Franchini2016},
within which these results where obtained. The interested reader may
look at the Sections 2 and 3 of \cite{FranchiniRSBwR2023}, Section
4 of \cite{Bardella}, the Theory Insights of \cite{BFPF}, and the
preprints \cite{Franchini2017,Franchini2016} for a comprehensive
presentation. 

\subsection*{Notation and fundamentals}

Let us introduce the notation. The set of vertices on which the spins
are placed will be denoted by $V$, with the vertices tracked by the
integer label $i$
\begin{equation}
V:=\{1\leq i\leq N\},
\end{equation}
the total number of vertices will be denoted by $N$, which from here
on is assumed to be a thermodynamically large integer. The binary
support of a single spin is
\begin{equation}
\Omega:=\{-1,1\}
\end{equation}
The generic spin state (magnetisation state) is as follows
\begin{equation}
\sigma_{V}:=\{\sigma_{i}\in\Omega:\,i\in V\}\in\Omega^{V}
\end{equation}
and represents one of the possible configurations of the system. Since
the support of the single spin is limited to two states, the cardinality
of the support of the magnetisation states will be equal to $|\Omega|=2$
raised to the number of vertices $N$,
\begin{equation}
\left|\Omega^{V}\right|=2^{N}
\end{equation}
in general, when we apply the absolute value to a set we mean its
cardinality. In what follows we will consider fixed the order $1\leq i\leq N$
in which the spins are labelled, this is reflected in the nonstandard
notation of the support 
\begin{equation}
\Omega^{\{1\}}\,\Omega^{\{2\}}\ ...\ \ \Omega^{\{N\}}=\prod_{i\in V}\ \Omega^{\{i\}}=:\Omega^{V}
\end{equation}
where, instead of simply raising to the power $N$ the generic binary
support $\Omega$, here we also want to keep track of the order in
which the individual spin 
\begin{equation}
\sigma_{i}\in\Omega^{\{i\}}
\end{equation}
are ordered to obtain the full magnetisation state $\sigma_{V}$.
It will be convenient to introduce a special notation for the uniform
measure: the uniform average will be indicated with the angled brackets
with the support in subscript
\begin{equation}
\langle\mathcal{O}\left(\sigma_{V}\right)\rangle_{\Omega^{V}}:=\frac{1}{|\,\Omega^{V}|}\sum_{\sigma_{V}\in\Omega^{V}}\mathcal{O}\left(\sigma_{V}\right)
\end{equation}
We also introduce the eigenstates of magnetisation, which we might
regard as the fundamental building blocks of our description. We call
``total magnetisation'' the operator that, if applied to a spin
state, returns its total magnetisation. The ``eigenstates of magnetisation''
with eigenvalue $M$ are therefore those spin states with magnetisation
exactly equal to $M$. Then, let us introduce the magnetisation function
\begin{equation}
M(\sigma_{V}):=\sum_{i\in V}\sigma_{i},\ \ \ m(\sigma_{V}):=\frac{M(\sigma_{V})}{N},
\end{equation}
that tracks the magnetisation of the spin vector. The value of the
magnetisation varies from the maximum $N$ to the minimum $-N$, corresponding
to all spin up and all spin down, respectively. Labelling with this
parameter we can further partition the support of $\sigma_{V}$ into
disjoint subsets of given magnetisation:
\begin{equation}
\Omega^{V}\left(M\right):=\{\sigma_{V}\in\Omega^{V}:\,M(\sigma_{V})=M\}
\end{equation}
the set collects all states of size $N$ and total magnetisation $M$.
The complete support is obtained by joining together the supports
with given magnetisation
\begin{equation}
\Omega^{V}=\bigcup_{M}\ \Omega^{V}\left(M\right)
\end{equation}
As for the case of the uniform average, we will use the angle brackets
(with the support in subscript) to indicate the average with respect
to the magnetisation eigenstates, 
\begin{equation}
\langle\mathcal{O}\left(\sigma_{V}\right)\rangle_{\,\Omega^{V}\left(M\right)}:=\frac{1}{|\,\Omega^{V}\left(M\right)|}\sum_{\sigma_{V}\in\Omega^{V}\left(M\right)}\mathcal{O}\left(\sigma_{V}\right)
\end{equation}
Magnetisation eigenstates exist also in the thermodynamic limit. To
see this, let
\begin{equation}
M=\left\lfloor mN\right\rfloor ,\ \ \ m\in\left[-1,1\,\right].
\end{equation}
From the theory of large deviations it is possible to deduce various
properties of eigenstates in the continuous limit as the magnetisation
parameter $m$ varies. To lighten the notation we introduce the following
abbreviations.
\begin{equation}
\Omega^{V}\left(m\right):=\Omega^{V}\left(\left\lfloor mN\right\rfloor \right)
\end{equation}
The average over the eigenstates with parameter $\left\lfloor mN\right\rfloor $
will be denoted simply by
\begin{equation}
\langle\mathcal{O}\left(\sigma_{V}\right)\rangle_{m}:=\langle\mathcal{O}\left(\sigma_{V}\right)\rangle_{\,\Omega^{V}\left(\left\lfloor mN\right\rfloor \right)}
\end{equation}
Since the eigenstates of the magnetisation are an ergodic set, the
average on the eigenstates can be computed numerically by generating
samples with a given magnetisation, evaluating the function and ultimately
averaging. Given the binary nature of the magnetisation values for
the single spin, it is also possible to obtain a detailed description
in the thermodynamic limit at the sample--path level (sample--path
Large Deviations) by applying the Varadhan lemma and the Mogulskii
theorem. For the teaching purpose of this paper we will not go into
these more advanced methods, but the interested reader may consult
\cite{FranchiniPhD2015,FranchiniURNS2017,FranchiniBalzanIRT2023}
and references therein for more details. 

\section{The general model\protect\label{sec:The-general-model}}

The general model we are going to study is a simple binary field theory,
truncated at second-order. For simplicity let us consider an isolated
system (with no external field), given an arbitrary interaction matrix
\begin{equation}
H=\{H_{ij}\in\Gamma:\,ij\in V^{2}\}\in\Gamma^{\,V^{2}}
\end{equation}
where the elementary support $\Gamma$ can be a finite alphabet or
interval or directly the real axis. The Hamiltonian of the model is
in general
\begin{equation}
H\left(\sigma_{V}\right):=\sum_{i\in V}\sum_{j\in V}H_{ij}\sigma_{i}\sigma_{j}
\end{equation}
this can be rewritten in a way that is convenient for us by introducing
the ``local fields'', or ``cavity fields'', depending on the context.
By grouping the local fields into a vector 
\begin{equation}
h_{V}\left(\sigma_{V}\right):=\{h_{i}\left(\sigma_{V}\right)\in\mathbb{R}:\,i\in V\},\ \ \ h_{i}\left(\sigma_{V}\right):=\sum_{j\in V}H_{ij}\sigma_{j}
\end{equation}
the Hamiltonian is rewritten as product between the state and the
local fields
\begin{equation}
H\left(\sigma_{V}\right)=\sigma_{V}\cdot h_{V}\left(\sigma_{V}\right).
\end{equation}
The partition function and the associated Gibbs measure (softmax)
are respectively
\begin{equation}
Z:=\sum_{\sigma_{V}\in\Omega^{V}}\exp\left[-H\left(\sigma_{V}\right)\right],\ \ \ \mu^{*}\left(\sigma_{V}\right):=\frac{\exp\left[-H\left(\sigma_{V}\right)\right]}{Z}
\end{equation}
we will use the symbol $\sim$ to indicate that a variable is distributed
according to a certain measure, and the symbol $\mathcal{P}$ to indicate
the probability space generated by the support within parentheses,
for example we will write
\begin{equation}
\sigma_{V}\sim\mu\in\mathcal{P}(\Omega^{V})
\end{equation}
to indicate that the magnetisation state is distributed according
to the chosen measure $\mu$, which belongs to the set of possible
probability measures contained in the support $\Omega^{V}$. The average
with respect to the generic probability measure will be indicated
with angled parenthesis with the measure indicated in subscript, for
example:
\begin{equation}
\langle\mathcal{O}\left(\sigma_{V}\right)\rangle_{\mu}:=\sum_{\sigma_{V}\in\Omega^{V}}\mu\left(\sigma_{V}\right)\mathcal{O}\left(\sigma_{V}\right).
\end{equation}
Apart from the case of uniform measure and magnetisation eigenstates,
this will be our standard notation for averaging over a given measurement. 

\subsection{Thermodynamic limit}

\noindent In canonical statistical theory, the thermodynamics of the
model is entirely described by the Helmholtz free energy function
\begin{equation}
f:=-\frac{1}{N}\log Z
\end{equation}
The very existence of the thermodynamic limit can in fact be reduced
to the existence of the limit of $f$ when $N\rightarrow\infty$,
which depends critically on how $H_{ij}$ scales in the spin number.
The most general form can be site--dependent, but for simplicity
we consider a smaller class in which the normalisation is uniform:
\begin{equation}
H_{ij}=\frac{\beta J_{ij}}{g\left(V^{2}\right)}
\end{equation}
here $J_{ij}$ are the coupling parameters and $g(V^{2})$ is a normalisation
that ensures the existence of $f$ in the thermodynamic limit, ie.
such that the components of $h_{V}$ converges to finite numbers (at
least in distribution). For a fully connected model this is usually
diverging in the spin number, and not dependent from the specific
labels of the spins. For example in the fully connected Curie-Weiss
model we find that the correct normalisation must scale with $N$,
while for the version with random weights, that is the SK model, the
correct normalisation is instead $\sqrt{N}$. Notice that to lighten
the notation we reabsorbed also the temperature in the parameters
$H_{ij}$.

\subsection{Free energy principle}

\noindent Applying the Jensen inequality we can easily deduce the
canonical free--energy principle \cite{Friston,Zeidman,Bardella}.
Let define the free energy functional
\begin{equation}
\mathscr{F}\left(\xi\right):=\langle H\left(\sigma_{V}\right)\rangle_{\xi}+\langle\,\log\xi\left(\sigma_{V}\right)\rangle_{\xi}
\end{equation}
and consider the following chain of inequalities \cite{FranchiniRSBwR2023,Zeidman}
\begin{multline}
\sum_{\sigma_{V}\in\Omega^{V}}\exp\left[-H\left(\sigma_{V}\right)\right]=\sum_{\sigma_{V}\in\Omega^{V}}\xi\left(\sigma_{V}\right)\exp\left[-H\left(\sigma_{V}\right)-\log\xi\left(\sigma_{V}\right)\right]=\\
=\langle\exp\left[-H\left(\sigma_{V}\right)-\log\xi\left(\sigma_{V}\right)\right]\rangle_{\xi}\geq\exp\left[-\mathscr{F}\left(\xi\right)\right]\label{eq:sgs}
\end{multline}
where in the last step we applied the Jensen inequality to the exponential
to bring the average inside. Then for any $\xi$ holds
\begin{equation}
Nf\leq\mathscr{F}\left(\xi\right),\ \ \ \forall\xi\in\mathscr{P}(\Omega^{V})
\end{equation}
It can be shown that the minimum is the Helmholtz free energy,
\begin{equation}
Nf=\inf_{\xi\in\mathscr{P}(\Omega^{V})}\mathscr{F}\left(\xi\right)
\end{equation}
that is attained by the Gibbs measure, $\mu^{*}$. This was the canonical
free energy principle. We can also deal with random Hamiltonians.
Let now suppose that the coupling matrix is distributed according
to some probability measure $\eta$
\begin{equation}
H\sim\eta\in\mathcal{P}(\Gamma^{\,V^{2}})
\end{equation}
and that we are interested in computing the quenched free energy,
\begin{equation}
\langle f\rangle_{\eta}:=\sum_{H\in\Gamma^{\,V^{2}}}\eta f
\end{equation}
Applying Jensen inequality again (to the logarithm this time) we find
the corresponding lower variational bound. From definitions we immediately
obtain 
\begin{multline}
-N\langle f\rangle_{\eta}=\sum_{H\in\Gamma^{\,V^{2}}}\eta\log Z=\sum_{H\in\Gamma^{\,V^{2}}}\log Z^{\,\eta}=\\
=\sum_{H\in\Gamma^{\,V^{2}}}\zeta\log Z^{\,\eta/\zeta}=\langle\log Z^{\,\eta/\zeta}\rangle_{\zeta},\ \ \ \forall\zeta\in\mathscr{P}(\Gamma^{\,V^{2}})\label{eq:vfvfvfvf}
\end{multline}
Let introduce the quenched free energy functional 
\begin{equation}
\mathscr{G}\left(\zeta\right):=-\log\,\langle\,Z^{\,\eta/\zeta}\rangle_{\zeta}=-\log\,\langle\exp\,\left(-N\eta f/\zeta\right)\rangle_{\zeta}
\end{equation}

\noindent By Jensen inequality holds the following
\begin{equation}
N\langle f\rangle_{\eta}\geq\mathscr{G}\left(\zeta\right),\ \ \ \forall\zeta\in\mathscr{P}(\Gamma^{\,V^{2}})
\end{equation}
from which we deduce a variational principle for the quenched free
energy
\begin{equation}
N\langle f\rangle_{\eta}=\sup_{\zeta\in\mathscr{P}(\Gamma^{\,V^{2}})}\mathscr{G}\left(\zeta\right)
\end{equation}
that is attained by the following measure 
\begin{equation}
\rho:=\frac{\eta f}{\langle f\rangle_{\eta}}
\end{equation}
this principle that takes the supremum of the functional instead of
the infimum is reminiscent of the Parisi variational principle for
the SK model or the Guerra interpolation. Notice that both principles
where obtained by inserting the identity operator in the form of either
$\log\,\exp$ or $\exp\log$ and then bringing the average one step
inside, 
\begin{equation}
\langle f\rangle=\langle\exp\,\log\left(f\right)\rangle\geq\exp\,\langle\log\left(f\right)\rangle,\ \ \ \langle f\rangle=\langle\log\,\exp\left(f\right)\rangle\leq\log\,\langle\exp\left(f\right)\rangle
\end{equation}
and should be ultimately equivalent. Notice also that combining with
the free energy principle we can introduce one last functional
\begin{multline}
-\mathscr{A}\left(\xi,\zeta\right):=\log\,\langle\exp\left[-\eta\mathscr{F}\left(\xi\right)/\zeta\right]\rangle_{\zeta}=\\
=\log\,\langle\exp\,[-\eta\langle H\rangle_{\xi}/\zeta-\eta\langle\log\xi\rangle_{\xi}/\zeta]\rangle_{\zeta}\label{eq:vfvfvfvf-2-1}
\end{multline}
that allows to write the free energy in terms of a min--max principle
\begin{equation}
\mathscr{G}\left(\zeta\right)=\inf_{\xi\in\mathscr{P}(\Omega^{V})}\mathscr{A}\left(\xi,\zeta\right)
\end{equation}
This is reminiscent of methods introduced by Guerra et al. and could
be related to the min-max principle identified by these authors for
the bipartite SK model \cite{Barra}. 

\section{Blanket (or Layer) representation\protect\label{sec:Blankets-(or-Layers)}}

Now that we have introduced the fundamental quantities, we can summarise
the method proposed in \cite{FranchiniSPA2021,FranchiniSPA2023}.
The theory builds on several Graph--Theoretical methods for statistical
physics and probability theory introduced by Lovaz, Borgs, Chayes,
Coja-Oghlan, and others, \cite{Borgs,Lovasz,Coja-Oghlan,Franchini2017,Franchini2016}.
The idea can be understood also in terms of partitioning the graph
into a progression of Markov blankets \cite{Judea_Pearl,Friston}
(that in \cite{FranchiniSPA2021,FranchiniSPA2023,FranchiniRSBwR2023}
where called ``layers''). For now we will not specify the detailed
structure of the blankets, which in general depends on the model,
we just discuss the general features. Then, let introduce a partition
of $V$ into $L$ disjoint subsets, hereafter recalled by the symbol
$\mathscr{V}$, 
\begin{equation}
V=\left\{ V_{\ell}\in\mathscr{V}:\,\ell\in\Lambda\right\} ,\ \ \ \Lambda:=\left\{ 1\leq\ell\leq L\right\} .
\end{equation}
As we shall see in short, the choice of the partition will be our
general order parameter. To facilitate comparison with other methods
it will be convenient to introduce a nested sequence of sets constructed
by progressively bringing the parts together following the order established
by the index $\ell$, then we introduce
\begin{equation}
Q_{\ell}:=\bigcup_{\ell^{*}\leq\ell}V_{\ell^{*}},\ \ \ Q_{\ell-1}\subseteq Q_{\ell},\ \ \ Q_{L}=V,
\end{equation}
This sequence is completely determined by the partition $\mathscr{V}$,
\begin{equation}
V_{\ell}:=Q_{\ell}\setminus Q_{\ell-1}.
\end{equation}
As we shall see, for densely connected models like the SK model only
the relative size of the parts will matter, then we introduce the
functional order parameter 
\begin{equation}
q:=\left\{ q_{\ell}\in\left[0,1\right]:\,q_{0}=0,\,q_{L}=1,\,q_{\ell}\leq q_{\ell+1},\,\ell\in\Lambda\right\} 
\end{equation}
that is a non--decreasing step function of $L$ steps bounded between
zero and one. The definition implies that $q$ is the quantile function
\cite{Steinbrecher} of some atomic probability measure. In fact,
the individual atoms, or steps 
\begin{equation}
p_{\ell}:=q_{\ell}-q_{\ell-1}\in\left[0,1\right]
\end{equation}
can be actually interpreted as probabilities. Then, the order parameter
is also uniquely associated to an atomic probability distribution
\begin{equation}
p:=\left\{ p_{\ell}\in\left[0,1\right]:\,\ell\in\Lambda\right\} \in\mathscr{P}(\Lambda)
\end{equation}
and we could equivalently consider the distribution $p$ as order
parameter. The relation between the size of the parts and the number
of spins is
\begin{equation}
|Q_{\ell}|=Nq_{\ell},\ \ \ |V_{\ell}|=N\left(q_{\ell}-q_{\ell-1}\right)=Np_{\ell}.
\end{equation}
Notice that the partition of $V$ induces a corresponding partition
$\mathscr{W}\left(\mathscr{V}\right)$ in the edges set $V^{2}$ that
allows to chart the energy contributions in the Hamiltonian. Let introduce
\begin{equation}
W_{\ell}:=Q_{\ell}^{2}\setminus Q_{\ell-1}^{2},
\end{equation}
that contains all the edges with both ends in $Q_{\ell}$ minus those
with both ends in $Q_{\ell-1}$. The cardinalities of these sets are
linked to the order parameter by 
\begin{equation}
|W_{\ell}|=|Q_{\ell}^{2}|-|Q_{\ell-1}^{2}|=N^{2}(\,q_{\ell}^{2}-q_{\ell-1}^{2})=N^{2}(p_{\ell}^{2}+2\,q_{\ell-1}p_{\ell})\label{eq:W}
\end{equation}
These coefficients are known to Spin Glass theory, since they appear
in the Parisi functional for the SK model. Notice that $p$ is the
discrete canonical conjugate of $q$, and that $p_{\ell}^{2}$ is
an infinitesimal quantity that can be ignored in the large $L$ limit.
Then, in the large $L$ limit the coefficient converges to the derivative
of $q_{\ell}^{2}$ respect to $\ell$.

\begin{figure}
\begin{centering}
\includegraphics[scale=0.16]{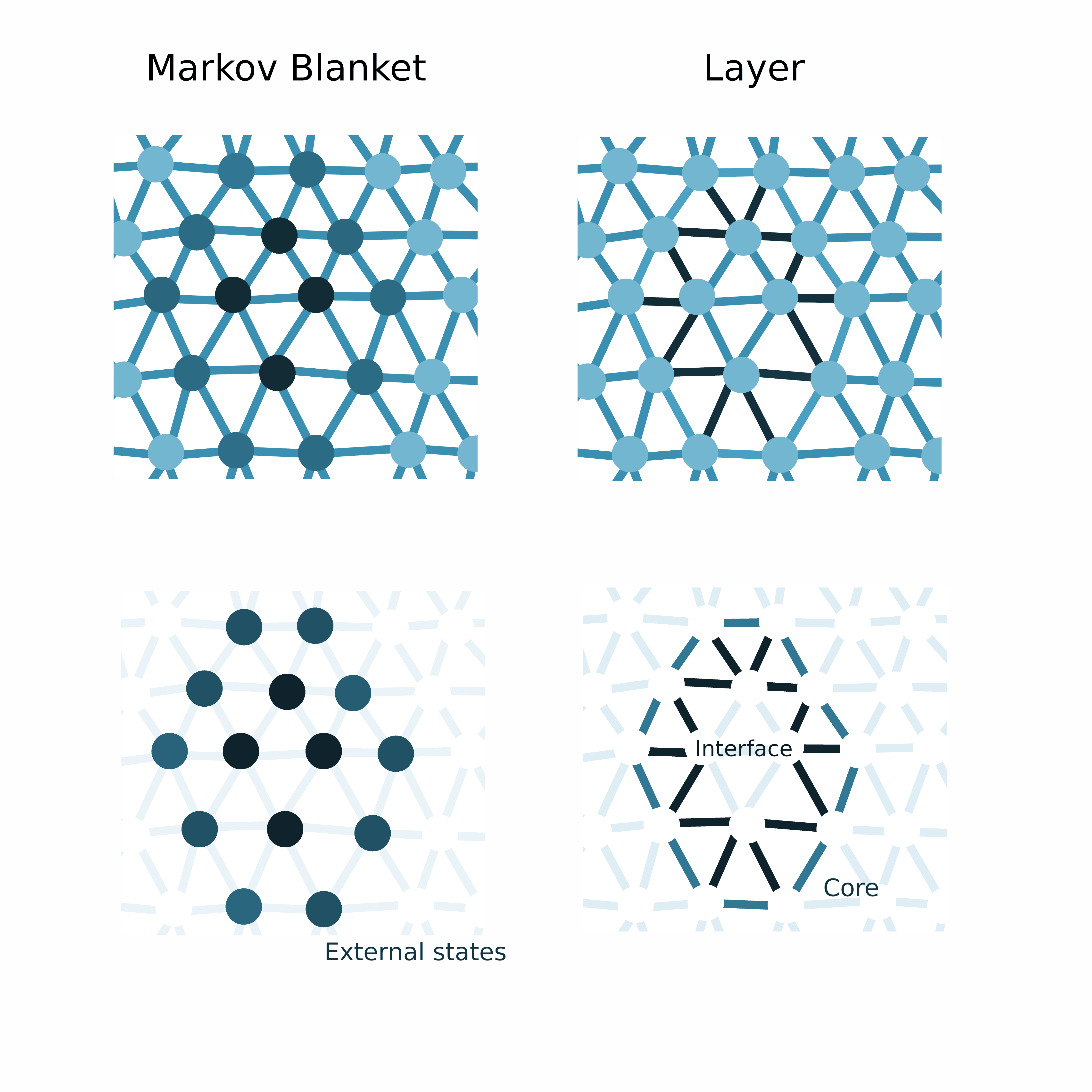}
\par\end{centering}
\begin{doublespace}
\caption{A comparison between the notion of a Markov blanket as is usually
intended (e.g., by Pearl \cite{Judea_Pearl} and Friston \cite{Friston})
versus the \textquotedblleft layer\textquotedblright{} representation
of \cite{FranchiniSPA2021,FranchiniSPA2023,FranchiniRSBwR2023}. As
one can see from the figures, the second case is just the interaction
picture of the first: the blanket is made of vertices while the layer
is made of edges. For this paper we will not make the distinction
and refer to both as Markov blankets. This is done to facilitate the
understanding with other fields that already use this concept.}
\end{doublespace}
\end{figure}

\subsection{Blankets and conditional expectations}

We can now apply the partition to the energy function and see if this
help in some way. Doing this, in \cite{FranchiniRSBwR2023} we identify
the Hamiltonian 

\begin{equation}
H_{\ell}\left(\sigma_{V_{\ell}}|\sigma_{Q_{\ell-1}}\right):=\sum_{\left(i,j\right)\in W_{\ell}}H_{ij}\sigma_{i}\sigma_{j}
\end{equation}
as the energy function that describes the particular blanket $\ell$.
The full Hamiltonian is recovered by simply summing the blankets
\begin{equation}
H\left(\sigma_{V}\right)=\sum_{\ell\leq L}H_{\ell}\left(\sigma_{V_{\ell}}|\sigma_{Q_{\ell-1}}\right)
\end{equation}
For each blanket we can associate the partition function
\begin{equation}
Z_{\ell}\left(\sigma_{Q_{\ell-1}}\right):=\sum_{\sigma_{V_{\ell}}\in\Omega^{V_{\ell}}}\exp\left[-H_{\ell}\left(\sigma_{V_{\ell}}|\sigma_{Q_{\ell-1}}\right)\right]
\end{equation}
and the associated Gibbs measure of the blanket
\begin{equation}
\xi_{\ell}\left(\sigma_{V_{\ell}}|\sigma_{Q_{\ell-1}}\right):=\frac{\exp\left[-H_{\ell}\left(\sigma_{V_{\ell}}|\sigma_{Q_{\ell-1}}\right)\right]}{Z_{\ell}\left(\sigma_{Q_{\ell-1}}\right)}
\end{equation}
The full probability distribution and the free energy functional are
obtained by multiplying the Gibbs measures of the individual blankets,
\begin{equation}
\mu\left(\sigma_{V}\right):=\prod_{\ell\leq L}\xi_{\ell}\left(\sigma_{V_{\ell}}|\sigma_{Q_{\ell-1}}\right),\ \ \ Z\left(\sigma_{V}\right):=\prod_{\ell\leq L}Z_{\ell}\left(\sigma_{Q_{\ell-1}}\right)\label{eq:fundamential}
\end{equation}
In particular, the second functional in the previous equation is related
to the partition function by the following fundamental formula:
\begin{equation}
Z=\langle Z\left(\sigma_{V}\right)\rangle_{\mu},\ \ \ \forall p\in\mathscr{P}(\Lambda)
\end{equation}
that expresses the flatness of the functional in Eq. (\ref{eq:fundamential})
respect to variations of the order parameter $p$. This result can
be shown easily by the following chain of identities:
\begin{multline}
Z=\sum_{\sigma_{V_{1}}\in\Omega^{V_{1}}}\,...\,\sum_{\sigma_{V_{L}}\in\Omega^{V_{L}}}\ \prod_{\ell\leq L}\ \exp\,[-H_{\ell}\left(\sigma_{V_{\ell}}|\sigma_{Q_{\ell-1}}\right)]=\\
=\sum_{\sigma_{V_{1}}\in\Omega^{V_{1}}}\exp\left[-H_{1}\left(\sigma_{V_{1}}\right)\right]\ \ \ ...\sum_{\sigma_{V_{L}}\in\Omega^{V_{L}}}\exp\left[-H_{L}\left(\sigma_{V_{L}}|\sigma_{Q_{L-1}}\right)\right]=\\
=\sum_{\sigma_{V_{1}}\in\Omega^{V_{1}}}Z_{1}\,\xi_{1}\left(\sigma_{V_{1}}\right)\ \ \ ...\sum_{\sigma_{V_{L}}\in\Omega^{V_{L}}}Z_{L}\left(\sigma_{Q_{L-1}}\right)\,\xi_{L}\left(\sigma_{V_{L}}|\sigma_{Q_{L-1}}\right)=\\
=\sum_{\sigma_{V_{1}}\in\Omega^{V_{1}}}\xi_{1}\left(\sigma_{V_{1}}\right)\ \ \ ...\sum_{\sigma_{V_{L-1}}\in\Omega^{V_{L-1}}}\xi_{L-1}\left(\sigma_{V_{L-1}}|\sigma_{Q_{L-2}}\right)\prod_{\ell\leq L}Z_{\ell}\left(\sigma_{Q_{\ell-1}}\right)=\\
=\sum_{\sigma_{V_{1}}\in\Omega^{V_{1}}}\ \ \ ...\sum_{\sigma_{V_{L-1}}\in\Omega^{V_{L-1}}}\,\prod_{\ell<L}\xi_{\ell}\left(\sigma_{V_{\ell}}|\sigma_{Q_{\ell-1}}\right)\prod_{\ell\leq L}Z_{\ell}\left(\sigma_{Q_{\ell-1}}\right)=\\
=\sum_{\sigma_{Q_{L-1}}\in\Omega^{Q_{L-1}}}\,\prod_{\ell<L}\xi_{\ell}\left(\sigma_{V_{\ell}}|\sigma_{Q_{\ell-1}}\right)\prod_{\ell\leq L}Z_{\ell}\left(\sigma_{Q_{\ell-1}}\right)=\\
=\langle\prod_{\ell\leq L}Z_{\ell}\left(\sigma_{Q_{\ell-1}}\right)\rangle_{\mu}=\langle Z\left(\sigma_{V}\right)\rangle_{\mu}.\label{eq:xffjsdtj}
\end{multline}
In general, the average of any observable $\mathcal{O}$ according
to $\mu^{*}$ is found trough a cascade application of conditional
expectations, i.e. by the following recursive formula:
\begin{equation}
\mathcal{O}_{L}\left(\sigma_{V}\right):=\mathcal{O}\left(\sigma_{V}\right),\ \ \ \mathcal{O}_{\ell-1}\left(\sigma_{Q_{\ell-1}}\right):=\langle\mathcal{O}_{\ell}\left(\sigma_{Q_{\ell}}\right)\rangle_{\xi_{\ell}},\ \ \ \mathcal{O}_{0}=\langle\mathcal{O}\left(\sigma_{V}\right)\rangle_{\mu}.\label{eq:therecursion}
\end{equation}
We iterate the recursion from the initial condition $\mathcal{O}_{L}$
until the last step $\mathcal{O}_{0}$ that correspond to the thermal
average of $\mathcal{O}$ according to $\mu$. Introducing the functional
\begin{equation}
f\left(\sigma_{V}\right):=-\frac{1}{N}\log Z\left(\sigma_{V}\right)
\end{equation}
that is a flat functional of the order parameter $p$, we can apply
the formula before 
\begin{equation}
f\left(\sigma_{V}\right)=\sum_{\ell\leq L}p_{\ell}f_{\ell}\left(\sigma_{Q_{\ell-1}}\right),\ \ \ f_{\ell}\left(\sigma_{Q_{\ell-1}}\right):=-\frac{1}{|V_{\ell}|}\log Z_{\ell}\left(\sigma_{Q_{\ell-1}}\right)
\end{equation}
and write the actual free energy in terms of the flat functional 
\begin{equation}
Nf=-\log\,\langle\,\exp\,[-Nf\left(\sigma_{V}\right)]\,\rangle_{\mu},\ \ \ \forall p\in\mathscr{P}(\Lambda).
\end{equation}
The usual free energy principle is recovered, as before, by bringing
the average with respect to $\mu$ inside the exponential, introducing
the functional
\begin{equation}
\mathcal{F}^{*}\left(p\right):=N\langle\,f\left(\sigma_{V}\right)\rangle_{\mu}
\end{equation}
we obtain the upper bound to the free energy:
\begin{equation}
Nf\leq\inf_{p\,\in\mathscr{P}(\Lambda)}\mathcal{F}^{*}
\end{equation}
Similarly, for a quenched free energy functional
\begin{equation}
N\langle f\rangle_{\eta}=-\langle\,\log\,\langle\,\exp\,[-Nf\left(\sigma_{V}\right)]\rangle_{\mu}\rangle_{\eta},\ \ \ \forall p\in\mathscr{P}(\Lambda).
\end{equation}
We can bring the average with respect to $p$ outside the exponential
\begin{equation}
\mathscr{G}^{*}\left(p\right):=-\langle\,\log\,\langle\,\langle\,\exp\,[-Nf_{\ell}\left(\sigma_{Q_{\ell-1}}\right)]\rangle_{\mu}\rangle_{p}\rangle_{\eta}
\end{equation}
and find a lower bound to the quenched free energy:
\begin{equation}
N\langle f\rangle_{\eta}\geq\sup_{p\,\in\mathscr{P}(\Lambda)}\mathscr{G}^{*}\left(p\right)
\end{equation}
We expect that this second variational principle is of the same kind
of that appearing in the celebrated Parisi formula for the SK model
\cite{RSB} where, in fact, the prescription is to maximise the energy
functional respect to the order parameter.

\subsection{Blanket internal structure}

It has been shown in Ref. \cite{FranchiniSPA2021} that the blankets
posses an internal structure. As one can easily appreciate from Figure
1 of \cite{FranchiniSPA2021}, it is possible to identify at least
two main types of energy contributions, those corresponding to edges
with both ends in $V_{\ell}$, that we call the core, and those with
one edge in $V_{\ell}$ and one in $Q_{\ell-1}$, that we call interface:
\begin{equation}
W_{\ell}=\hat{W}_{\ell}\cup\bar{W}_{\ell},\ \ \ \hat{W}_{\ell}:=V_{\ell}^{2},\ \ \ \bar{W}_{\ell}:=\left(V_{\ell}\,Q_{\ell-1}\right)\cup\left(Q_{\ell-1}V_{\ell}\right)
\end{equation}
Hereafter, we indicate with a hat on top if the quantity is relative
to the core and with a bar if to the interface. This partition is
also shown in Figures 4.1 and 4.2 of \cite{FranchiniRSBwR2023}. The
effect on the Hamiltonian of this further partition is best understood
by introducing the normalisation coefficients
\begin{equation}
\hat{\gamma}_{\ell}:=g\left(V_{\ell}^{2}\right)/g\left(V^{2}\right),\ \ \ \bar{\gamma}_{\ell}:=g\left(\bar{W}_{\ell}\right)/g\left(V^{2}\right)
\end{equation}
Using these coefficients, that for densely connected models can be
non--trivial, due to the dependence of $g(V^{2})$ on the number
of spins, the core of the blanket (or layer) is
\begin{equation}
\hat{H}_{\ell}\left(\sigma_{V_{\ell}}\right):=\hat{\gamma}_{\ell}H\left(\sigma_{V_{\ell}}\right)
\end{equation}
and is simply a smaller model (multiplied by a normalisation). The
interface is instead described by a very simple bipartite Hamiltonian,
that is the scalar product between the spins of the blanket and the
local fields:
\begin{equation}
\bar{H}_{\ell}\left(\sigma_{V_{\ell}}|\sigma_{Q_{\ell-1}}\right):=\sigma_{V_{\ell}}\cdot\bar{\gamma}_{\ell}h_{V_{\ell}}\left(\sigma_{Q_{\ell-1}}\right)
\end{equation}
The full blanket is then recovered by summing back these contributions
\begin{equation}
H_{\ell}\left(\sigma_{V_{\ell}}|\sigma_{Q_{\ell-1}}\right)=\hat{H}_{\ell}\left(\sigma_{V_{\ell}}\right)+\bar{H}_{\ell}\left(\sigma_{V_{\ell}}|\sigma_{Q_{\ell-1}}\right)
\end{equation}
Notice that if we interpret $\ell$ as a sort of discrete time then
the blanket structure is reminiscent of the celebrated Transformer
architecture \cite{Google}, with the cores corresponding to the attention
modules \cite{Rendt2024}. Let conclude by studying the effects of
this decomposition on the partition function: to properly explain
this it will be convenient to introduce the partition function of
the interface,
\begin{equation}
\bar{Z}_{\ell}\left(\sigma_{Q_{\ell-1}}\right):=\prod_{i\in V_{\ell}}2\cosh\left[\bar{\gamma}_{\ell}h_{i}\left(\sigma_{Q_{\ell-1}}\right)\right]
\end{equation}
and the associated Gibbs measure (softmax)
\begin{equation}
\bar{\xi}_{\ell}\left(\sigma_{V_{\ell}}|\sigma_{Q_{\ell-1}}\right):=\frac{\exp\left[-\sigma_{V_{\ell}}\cdot\bar{\gamma}_{\ell}h_{V_{\ell}}\left(\sigma_{Q_{\ell-1}}\right)\right]}{\bar{Z}_{\ell}\left(\sigma_{Q_{\ell-1}}\right)}
\end{equation}
Then, the average with respect to the interface is given by the formula
\cite{FranchiniSPA2023}
\begin{equation}
\langle\mathcal{O}\left(\sigma_{V_{\ell}}\right)\rangle_{\bar{\xi}_{\ell}}=\frac{1}{\bar{Z}_{\ell}\left(\sigma_{Q_{\ell-1}}\right)}\sum_{\sigma_{V_{\ell}}\in\Omega^{V_{\ell}}}\mathcal{O}\left(\sigma_{V_{\ell}}\right)\,\exp\left[-\sigma_{V_{\ell}}\cdot\bar{\gamma}_{\ell}h_{V_{\ell}}\left(\sigma_{Q_{\ell-1}}\right)\right]
\end{equation}
Concerning the core, the partition function is just a number as all
spin degrees of freedom gets integrated out, the Gibbs average is
then just a smaller model restricted to the spins of the blanket and
rescaled by a factor $\hat{\gamma}_{\ell}$
\begin{equation}
\hat{Z}_{\ell}:=\sum_{\sigma_{V_{\ell}}\in\Omega^{V_{\ell}}}\exp\,[-\hat{\gamma}_{\ell}\,H\left(\sigma_{V_{\ell}}\right)],\ \ \ \hat{\xi}_{\ell}\left(\sigma_{V_{\ell}}\right):=\frac{\exp\,[-\hat{\gamma}_{\ell}H\left(\sigma_{V_{\ell}}\right)]}{\hat{Z}_{\ell}}
\end{equation}
the average is indicated like before 
\begin{equation}
\langle\mathcal{O}\left(\sigma_{V_{\ell}}\right)\rangle_{\hat{\xi}_{\ell}}=\frac{1}{\hat{Z}_{\ell}}\sum_{\sigma_{V_{\ell}}\in\Omega^{V_{\ell}}}\mathcal{O}\left(\sigma_{V_{\ell}}\right)\,\exp\,[-\hat{\gamma}_{\ell}H\left(\sigma_{V_{\ell}}\right)]
\end{equation}
Let compute the full blanket average. We start from the identity
\begin{equation}
\exp\,[-H_{\ell}\left(\sigma_{V_{\ell}}|\sigma_{Q_{\ell-1}}\right)]=\exp\,[-\hat{\gamma}_{\ell}H\left(\sigma_{V_{\ell}}\right)]\,\exp\,[-\sigma_{V_{\ell}}\cdot\bar{\gamma}_{\ell}h_{V_{\ell}}\left(\sigma_{Q_{\ell-1}}\right)]
\end{equation}
and integrate the spin variables in two steps. Notice that the order
in which the integrations are performed, although not influent on
the final result, can give rise to different (equivalent) expressions.
For example, if we integrate the core first 
\begin{multline}
\langle\mathcal{O}\left(\sigma_{V_{\ell}}\right)\rangle_{\xi_{\ell}}=\sum_{\sigma_{V_{\ell}}\in\Omega^{V_{\ell}}}\hat{\xi}_{\ell}\left(\sigma_{V_{\ell}}\right)\frac{\mathcal{O}\left(\sigma_{V_{\ell}}\right)\,\exp\,[-\sigma_{V_{\ell}}\cdot\bar{\gamma}_{\ell}h_{V_{\ell}}\left(\sigma_{Q_{\ell-1}}\right)]}{\langle\exp\,[-\sigma_{V_{\ell}}\cdot\bar{\gamma}_{\ell}h_{V_{\ell}}\left(\sigma_{Q_{\ell-1}}\right)]\rangle_{\hat{\xi}_{\ell}}}=\\
=\frac{\langle\mathcal{O}\left(\sigma_{V_{\ell}}\right)\,\exp\,[-\sigma_{V_{\ell}}\cdot\bar{\gamma}_{\ell}h_{V_{\ell}}\left(\sigma_{Q_{\ell-1}}\right)]\rangle_{\hat{\xi}_{\ell}}}{\langle\exp\,[-\sigma_{V_{\ell}}\cdot\bar{\gamma}_{\ell}h_{V_{\ell}}\left(\sigma_{Q_{\ell-1}}\right)]\rangle_{\hat{\xi}_{\ell}}}\label{eq:corefirst}
\end{multline}
while integrating the interface first we find
\begin{multline}
Z_{\ell}\left(\sigma_{Q_{\ell-1}}\right)=\sum_{\sigma_{V_{\ell}}\in\Omega^{V_{\ell}}}\exp\,[-H_{\ell}\left(\sigma_{V_{\ell}}|\sigma_{Q_{\ell-1}}\right)]=\\
=\sum_{\sigma_{V_{\ell}}\in\Omega^{V_{\ell}}}\exp\,[-\hat{\gamma}_{\ell}H\left(\sigma_{V_{\ell}}\right)]\,\exp\,[-\sigma_{V_{\ell}}\cdot\bar{\gamma}_{\ell}h_{V_{\ell}}\left(\sigma_{Q_{\ell-1}}\right)]=\\
=\bar{Z}_{\ell}\left(\sigma_{Q_{\ell-1}}\right)\sum_{\sigma_{V_{\ell}}\in\Omega^{V_{\ell}}}\bar{\xi}_{\ell}\left(\sigma_{V_{\ell}}|\sigma_{Q_{\ell-1}}\right)\exp\,[-\hat{\gamma}_{\ell}H\left(\sigma_{V_{\ell}}\right)]=\\
=\bar{Z}_{\ell}\left(\sigma_{Q_{\ell-1}}\right)\langle\exp\,[-\hat{\gamma}_{\ell}H\left(\sigma_{V_{\ell}}\right)]\rangle_{\bar{\xi}_{\ell}}\label{eq:fgfgf}
\end{multline}
for the partition function and 
\begin{multline}
\langle\mathcal{O}\left(\sigma_{V_{\ell}}\right)\rangle_{\xi_{\ell}}=\sum_{\sigma_{V_{\ell}}\in\Omega^{V_{\ell}}}\mathcal{O}\left(\sigma_{V_{\ell}}\right)\,\frac{\exp\,[-H_{\ell}\left(\sigma_{V_{\ell}}|\sigma_{Q_{\ell-1}}\right)]}{Z_{\ell}\left(\sigma_{Q_{\ell-1}}\right)}=\\
=\sum_{\sigma_{V_{\ell}}\in\Omega^{V_{\ell}}}\mathcal{O}\left(\sigma_{V_{\ell}}\right)\,\frac{\exp\,[-\hat{\gamma}_{\ell}H\left(\sigma_{V_{\ell}}\right)]\,\exp\,[-\sigma_{V_{\ell}}\cdot\bar{\gamma}_{\ell}h_{V_{\ell}}\left(\sigma_{Q_{\ell-1}}\right)]}{\bar{Z}_{\ell}\left(\sigma_{Q_{\ell-1}}\right)\langle\exp\,[-\hat{\gamma}_{\ell}H\left(\sigma_{V_{\ell}}\right)]\rangle_{\bar{\xi}_{\ell}}}=\\
=\sum_{\sigma_{V_{\ell}}\in\Omega^{V_{\ell}}}\bar{\xi}_{\ell}\left(\sigma_{V_{\ell}}|\sigma_{Q_{\ell-1}}\right)\frac{\mathcal{O}\left(\sigma_{V_{\ell}}\right)\,\exp\,[-\hat{\gamma}_{\ell}H\left(\sigma_{V_{\ell}}\right)]}{\langle\exp\,[-\hat{\gamma}_{\ell}H\left(\sigma_{V_{\ell}}\right)]\rangle_{\bar{\xi}_{\ell}}}=\\
=\frac{\langle\mathcal{O}\left(\sigma_{V_{\ell}}\right)\,\exp\,[-\hat{\gamma}_{\ell}H\left(\sigma_{V_{\ell}}\right)]\rangle_{\bar{\xi}_{\ell}}}{\langle\exp\,[-\hat{\gamma}_{\ell}H\left(\sigma_{V_{\ell}}\right)]\rangle_{\bar{\xi}_{\ell}}}\label{eq:interfacefirst}
\end{multline}
for the full blanket average. What is interesting in the first case
in Eq. (\ref{eq:corefirst}) (core first) is that the averaged quantities
all depend on the core distributions. This is the approach that was
actually taken in \cite{FranchiniSPA2021} to prove the dominance
of the interfaces in the SK model and may be convenient in some special
cases, but we now believe that the most useful expression is the second
case in Eq. (\ref{eq:interfacefirst}) (interface first), because
the averages are done respect to the ``interface'' model \cite{FranchiniSPA2023}
that is a one--body Hamiltonian, is universal and is generally much
easier to control, as is equivalent to adding an external field. 

\section{Fully connected models\protect\label{sec:Fully-connected-models}}

In \cite{FranchiniSPA2021,FranchiniSPA2023,FranchiniRSBwR2023} we
showed that in fully connected models with reasonable coupling strength
distribution the core can be ignored if the partition is fine enough. 

In general, for a finite dimensional model, and any other network
with irreducible core, we have that the ratio between the core and
the interface converges to a constant
\begin{equation}
\hat{\gamma}_{\ell}/\bar{\gamma}_{\ell}\rightarrow\mathrm{const.}
\end{equation}
hence both the cores and the interfaces contribute to the total energy
of the blankets in a comparable way also for an infinitely refined
partition, i.e., for $L\rightarrow\infty$. In this case the theory
is non--trivial and non--mean field. This class includes, e.g.,
the Ising model in two and three dimension, polymers \cite{FranchiniBalzanRANGE2018,FranchiniMS2011,Franchini2011},
random models like the Edaward-Anderson model, but also the Transformer
architecture by Google \cite{Google,Rendt2024}, the neural Lattice
Field Theory (LFT) in Section 4 of \cite{Bardella}, etc. 

Then, we have two types of mean field models: the first is what we
call the core-only (CO) model, is such that the interface can be ignored
\begin{equation}
\hat{\gamma}_{\ell}/\bar{\gamma}_{\ell}\rightarrow\infty,\ \ \ f\rightarrow\hat{f}
\end{equation}
and is the most typical among the mean--field approximations \cite{Opper}.
Practically, it consists in finding a partition (if any) in which
the interfaces can be neglected. In such way the measure is just the
product measure of the cores, the blankets are approximated as independent.

Finally, the second kind of mean--field approximation considered
in \cite{FranchiniSPA2021,FranchiniSPA2023,FranchiniRSBwR2023} is
the interface only (IO) model, where 
\begin{equation}
\hat{\gamma}_{\ell}/\bar{\gamma}_{\ell}\rightarrow0,\ \ \ f\rightarrow\bar{f}
\end{equation}
i.e., where the core contributions are overwhelmed by the interfaces,
or where we can identify the blanket sequence in such way that the
core is zero, e.g. like a purely feed--forward network or the levels
of some Caylay tree, if we want to consider a model with finite connectivity.
As we shall see, this class includes the Curie--Weiss model and the
SK model. 

\subsection{General theory of the IO model\protect\label{subsec:General-theory-of}}

Since the interface seems to posses a quite universal scaling limit,
let construct a general theory of the IO model \cite{FranchiniRSBwR2023}.
We will start from the IO Hamiltonian

\begin{equation}
\bar{H}\left(\sigma_{V}\right):=\sum_{\ell\in\Lambda}\sigma_{V_{\ell}}\cdot\bar{\gamma}_{\ell}h_{V_{\ell}}\left(\sigma_{Q_{\ell-1}}\right)
\end{equation}
If we arrange the field components in decreasing order we can further
split $V_{\ell}$ into $L'$ sub-blankets with approximately constant
value. We call this partition $\mathscr{V}_{\ell}$
\begin{equation}
V_{\ell}=\{V_{\ell\ell'}\in\mathscr{V}_{\ell}:\,\ell'\in\Lambda'\},\ \ \ \Lambda':=\{1\leq\ell'\leq L'\}
\end{equation}
For simplicity hereafter we take $L'=L$. We can identify two bounds,
both dependent on the previous blankets but not on sub-blankets of
the same blanket, and such that 
\begin{equation}
h_{\ell\ell'}^{-}\left(\sigma_{Q_{\ell-1}}\right)\leq h_{i}\left(\sigma_{Q_{\ell-1}}\right)\le h_{\ell\ell'}^{+}\left(\sigma_{Q_{\ell-1}}\right),\ \ \ \forall i\in V_{\ell\ell'}
\end{equation}
Most importantly, the gap shrinks proportionally to the inverse of
$L^{2}$ \cite{FranchiniRSBwR2023}
\begin{equation}
h_{\ell\ell'}^{+}\left(\sigma_{Q_{\ell-1}}\right)-h_{\ell\ell'}^{-}\left(\sigma_{Q_{\ell-1}}\right)=O\left(1/L^{2}\right)
\end{equation}
and is therefore converging to zero in the limit of infinite layers
(blankets). Introducing some useful additional notation
\begin{equation}
\hat{\gamma}_{\ell\ell'}:=g\left(V_{\ell\ell'}^{2}\right)/g\left(V^{2}\right),\ \ \ \hat{h}_{\ell\ell'}^{\pm}\left(\sigma_{Q_{\ell-1}}\right):=\hat{\gamma}_{\ell\ell'}h_{\ell\ell'}^{\pm}\left(\sigma_{Q_{\ell-1}}\right)
\end{equation}
we immediately see that the Hamiltonian of the IO model satisfies
\begin{equation}
\sum_{\ell\in\Lambda}\,\sum_{\ell'\in\Lambda'}\hat{h}_{\ell\ell'}^{-}(\sigma_{Q_{\ell-1}})\,m(\sigma_{V_{\ell\ell'}})\leq\frac{\bar{H}\left(\sigma_{V}\right)}{N}\leq\sum_{\ell\in\Lambda}\,\sum_{\ell'\in\Lambda'}\hat{h}_{\ell\ell'}^{+}(\sigma_{Q_{\ell-1}})\,m(\sigma_{V_{\ell\ell'}})\label{eq:Bound}
\end{equation}
and can be therefore approximated by energy functions that are supported
by the magnetisation eigenstates, similarly to what conjectured Mezard
and Virasoro in \cite{MezVir} for the SK model. This is important,
as it allows to work also in the thermodynamic limit using well known
methods \cite{Dembo}. Most importantly, we found that under Gibbs
measure the fields can be approximated by the parameter matrices
\begin{equation}
h\in\mathbb{R}^{\,L^{2}},\ \ \ m\in\left[-1,1\right]^{\,L^{2}}
\end{equation}
in the thermodynamic limit, and that the sub-blanket average
\begin{equation}
\lim_{N\rightarrow\infty}\langle\mathcal{O}\left(\sigma_{V_{\ell\ell'}}\right)\rangle_{\bar{\xi}_{\ell\ell'}}=\lim_{N\rightarrow\infty}\langle\mathcal{O}\left(\sigma_{V_{\ell\ell'}}\right)\rangle_{m_{\ell\ell'}}
\end{equation}
will converge to an average over magnetisation eigenstates with given
eigenvalue. This last step allows to actually compute the interface
averages numerically as function of the parameters by applying the
recursion in Eq. (\ref{eq:therecursion}) for each layer and sub--layer. 

Notice that the magnetisation eigenstates commute, i.e., the overlap
matrix and the correlation matrix both converge to the square of the
magnetisations (in the thermodynamic limit) \cite{FranchiniSPA2023}.
The implications of this are explained in Section 2 of \cite{FranchiniRSBwR2023}. 

Also, in \cite{FranchiniSPA2023,FranchiniRSBwR2023} we showed that,
at least in the small temperature limit, the IO model converges to
a Generalised Random Energy Model (GREM) of the Derrida's type \cite{Kurkova1,Kistler}
due to a general phenomenon known as REM universality \cite{Arous-Kupsov},
and that this result can be extended to any temperature at least in
case of Gaussian couplings \cite{FranchiniSPA2023}. 

Finally, notice that due to the new level of partitioning the order
parameter evolved from a vector into a matrix, but then, notice that
since there is no space structure only the levels matter, and the
order parameter is ultimately the distribution of the entries of the
parameter matrix, i.e., the quantile $h\left(m\right)$. 

\subsection{Free energy functional}

Now that we have a method to compute the averages, it only remains
to find a manageable expression for the free energy functional. We
start from

\begin{equation}
\bar{f}_{\ell}\left(\sigma_{Q_{\ell-1}}\right)=-\frac{1}{|V_{\ell}|}\sum_{i\in V_{\ell}}\log2\cosh\left[\bar{\gamma}_{\ell}\,h_{i}\left(\sigma_{Q_{\ell-1}}\right)\right]
\end{equation}
This is the normalised expression for the free energy density of the
blanket. Notice that for uniform models (i.e. when the order of the
spins don't matter) it is possible to make the parts of any size we
want: for example, taking the finest 
\begin{equation}
V_{i}=\left\{ i\right\} ,\ \ \ Q_{i}=\left\{ 1,\,2,\,...\,,\,i\,\right\} 
\end{equation}
the coefficient are like before
\begin{equation}
\hat{\gamma}_{i}:=g\left(V_{i}^{2}\right)/g\left(V^{2}\right),\ \ \ \bar{\gamma}_{i}:=g\left(\bar{W}_{i}\right)/g\left(V^{2}\right)
\end{equation}
but see the most elementary energy contribution is now composed by
just a single spin coupled to an element of the coupling matrix 
\begin{equation}
h_{i}(\sigma_{j}):=H_{ij}\sigma_{j}
\end{equation}
The cavity field is then rewritten as follows
\begin{equation}
\bar{\gamma}_{i}\,h_{i}(\sigma_{Q_{i-1}})=\sum_{j<i}\hat{\gamma}_{j}h_{i}(\sigma_{j})
\end{equation}
and by substituting in the definition we find 
\begin{equation}
\bar{f}_{i}\left(\sigma_{Q_{i-1}}\right)=-\log2\cosh\left[\bar{\gamma}_{i}\,h_{i}(\sigma_{Q_{i-1}})\right]=-\log2\cosh\,\sum_{j<i}\hat{\gamma}_{j}\,h_{i}(\sigma_{j}).
\end{equation}
In the thermodynamic limit the index is continuous \cite{FranchiniSPA2023},
let 
\begin{equation}
i/|V_{\ell}|\rightarrow x\in\left[0,1\right],\ \ \ j/|V_{\ell}|\rightarrow y\in\left[0,1\right]
\end{equation}
the coefficients and the cavity fields converge to functions
\begin{equation}
\bar{\gamma}_{i}\rightarrow\gamma\left(x\right),\ \ \ \hat{\gamma}_{j}\rightarrow d\gamma\left(y\right),\ \ \ h_{i}(\sigma_{j})\rightarrow h\left[\,x,\sigma\left(y\right)\right]
\end{equation}
putting together, the continuum limit of the interface is given by
the integral \cite{FranchiniSPA2023}
\begin{equation}
\bar{f}_{\ell}\left(\sigma_{Q_{\ell-1}}\right)\rightarrow-\int_{0}^{1}dx\,\log2\cosh\int_{0}^{x}d\gamma\left(y\right)\,h\left[\,x,\sigma\left(y\right)\right]
\end{equation}
then, if the field is non random we can simply apply the chain rule,
while for fields with random coefficients the expression converges
to an Ito integral, that can be analysed by Ito calculus, see for
example \cite{ChenAuffinger}. An alternative way to find the free
energy functional for the recursion of Eq. (\ref{eq:therecursion})
is the ``Cavity Method'' (CM), \cite{RSB}. The CM is shown in the
kernel context in the Section 6 of \cite{FranchiniRSBwR2023} for
the SK model, using a mathematical approach originally developed by
Aizenmann, Sims and Starr \cite{ASS} (see also \cite{FranchiniSPA2021}
or \cite{Franchini2019}).

\subsection{Example 1: Curie--Weiss model}

As first practical application we analyse the Curie-Weiss model \cite{CW},
a fully connected model where the interaction strength is constant.
The parameters are
\begin{equation}
H_{ij}=\beta J_{0}/N,\ \ \ g\left(V^{2}\right)=N
\end{equation}
Since the prescription for the existence of the thermodynamic limit
is that the components of the cavity fields stay finite as the system
grows, then, the cavity field must be normalised by a quantity that
is linear in the number of spin. Then we must have the following form
of the cavity field
\begin{equation}
h_{i}\left(\sigma_{V}\right)=\frac{J_{0}}{N}\sum_{j\in V}\sigma_{j}=J_{0}\,m\left(\sigma_{V}\right)
\end{equation}
and the full Hamiltonian is just proportional to the square of the
magnetisation operator
\begin{equation}
H_{\mathrm{CW}}\left(\sigma_{V}\right):=\frac{J_{0}}{N}\sum_{i\in V}\sum_{j\in V}\sigma_{i}\sigma_{j}=NJ_{0}\,m\left(\sigma_{V}\right)^{2}
\end{equation}
Most importantly, we can explicitly compute the coefficients 
\begin{equation}
\hat{\gamma}_{\ell}=p_{\ell},\ \ \ \bar{\gamma}_{\ell}=2q_{\ell-1},\ \ \ \hat{\gamma}_{\ell}/\bar{\gamma}_{\ell}=p_{\ell}/2q_{\ell-1}\rightarrow0
\end{equation}
and find, as predicted, that in the TL and infinite levels of partition
the core energy
\begin{equation}
\hat{H}_{\ell}\left(\sigma_{V_{\ell}}\right)=p_{\ell\,}H_{\mathrm{CW}}(\sigma_{V_{\ell}})=NJ_{0}\,p_{\ell}^{2}\,m\left(\sigma_{V_{\ell}}\right)^{2}
\end{equation}
becomes negligible in magnitude respect to the interface energy
\begin{equation}
\bar{H}_{\ell}\left(\sigma_{V_{\ell}}|\sigma_{Q_{\ell-1}}\right)=NJ_{0}\,2q_{\ell-1}m\left(\sigma_{Q_{\ell-1}}\right)=NJ_{0}\,\sum_{\ell^{*}<\ell}2p_{\ell^{*}}m\,(\sigma_{V_{\ell^{*}}})
\end{equation}
Notice in the last step we expanded the magnetisation states into
the local magnetisations: we used a different superscript for the
auxiliary variable $\ell^{*}$ to avoid confusion with the sub-index
$\ell^{'}$ before, that has quite different meaning. The free energy
functional of the blanket is then
\begin{multline}
\bar{f}_{\ell}\left(\sigma_{Q_{\ell-1}}\right)=-\frac{1}{\beta}\log2\cosh\,[\,\beta J_{0}\,2q_{\ell-1}m\,(\sigma_{Q_{\ell-1}})]=\\
=-\frac{1}{\beta}\log2\cosh\,\beta J_{0}\sum_{\ell^{*}<\ell}2p_{\ell^{*}}m\,(\sigma_{V_{\ell^{*}}})\label{eq:yufif}
\end{multline}
and we can explicitly compute the free energy density by applying
the cascade of conditional expectations described in Eq. (\ref{eq:therecursion}). 

\subsection{Example 2: Sherrington--Kirkpatrick model}

Let now move to the glassy version of the CW model, that is the SK
model. The model parameters are random this time
\begin{equation}
H_{ij}=\beta J_{ij}/\sqrt{N},\ \ \ J_{ij}\sim\mathcal{N}\left(0,1\right),\ \ \ g\left(V^{2}\right)=\sqrt{N}
\end{equation}
the normalisation is the square root of the number of spins
\begin{equation}
h_{i}\left(\sigma_{V}\right)=\frac{1}{\sqrt{N}}\sum_{j\in V}J_{ij}\sigma_{j}
\end{equation}
the Hamiltonian of the (asymmetric) SK model is
\begin{equation}
H_{\mathrm{SK}}\left(\sigma_{V}\right):=\frac{1}{\sqrt{N}}\sum_{i\in V}\sum_{j\in V}J_{ij}\sigma_{i}\sigma_{j}
\end{equation}
this immediately imply the following form of the coefficients
\begin{equation}
\hat{\gamma}_{\ell}=\sqrt{p_{\ell}},\ \ \ \bar{\gamma}_{\ell}=\sqrt{2q_{\ell-1}},\ \ \ \hat{\gamma}_{\ell}/\bar{\gamma}_{\ell}=\sqrt{p_{\ell}/2q_{\ell-1}}\rightarrow0
\end{equation}
and by confronting the expressions of the core vs interface energies
\begin{equation}
\hat{H}_{\ell}\left(\sigma_{V_{\ell}}\right)=\sqrt{p_{\ell}}\ H_{\mathrm{SK}}\left(\sigma_{V_{\ell}}\right),\ \ \ \bar{H}_{\ell}\left(\sigma_{V_{\ell}}|\sigma_{Q_{\ell-1}}\right)=\sigma_{V_{\ell}}\cdot\sqrt{2q_{\ell-1}}\ h_{V_{\ell}}\left(\sigma_{Q_{\ell-1}}\right)
\end{equation}
we can see that the interface contribution is still dominant in the
thermodynamic limit. Then, expanding the cavity field like before
\begin{equation}
\sqrt{2q_{\ell-1}}\ h_{i}(\sigma_{Q_{\ell-1}})=\sum_{\ell^{*}<\ell}\sqrt{2p_{\ell^{*}}}\,\,h_{i}(\sigma_{V_{\ell^{*}}})
\end{equation}
we arrive to the free energy functional of the SK model 
\begin{multline}
\bar{f}_{\ell}\left(\sigma_{Q_{\ell-1}}\right)=-\frac{1}{\beta|V_{\ell}|}\sum_{i\in V_{\ell}}\log2\cosh\,\beta\sqrt{2q_{\ell-1}}\,\,h_{i}(\sigma_{Q_{\ell-1}})=\\
=-\frac{1}{\beta|V_{\ell}|}\sum_{i\in V_{\ell}}\log2\cosh\,\beta\sum_{\ell^{*}<\ell}\sqrt{2p_{\ell^{*}}}\,\,h_{i}(\sigma_{V_{\ell^{*}}})\label{eq:fsadf}
\end{multline}
As said before, an alternative way to find the functional is the Cavity
Method, \cite{RSB,ASS}: we showed a detailed derivation for the SK
model in \cite{FranchiniSPA2021,FranchiniSPA2023,FranchiniRSBwR2023}
(see also in Ref. \cite{Franchini2019} for the shortest derivation).
Also, look at the Section 4 of \cite{Bardella}, in Paragraph 4.2.1,
for how to obtain the RSB theory as the special stationary case of
the kernel picture, within the larger context of Lattice Field Theories
\cite{FranchiniRSBwR2023,Bardella,BFPF,Franchini2017,Franchini2016}.

\section*{Acknowledgements}

We are grateful to Pan Liming (USTC), Giampiero Bardella (Sapienza
Università Ro\-ma), and Nicola Kistler (Goethe University Frankfurt)
for interesting discussions and suggestions. This research was partially
funded by European Research Council (ERC), under the European Union's
Horizon 2020 research and innovation programme (Grant Agreement No.
694925).

\end{document}